# Ripple modulated electronic structure of a 3D topological insulator

Yoshinori Okada, W. Zhou, D. Walkup, C. Dhital, S. D. Wilson, V. Madhavan

3D topological insulators, similar to the Dirac material graphene, host linearly dispersing states with unique properties and a strong potential for applications. A key, missing element in realizing some of the more exotic states in topological insulators is the ability to manipulate local electronic properties. Analogy with graphene suggests a possible avenue via a topographic route by the formation of superlattice structures such as a moiré patterns or ripples, which can induce controlled potential variations. However, while the charge and lattice degrees of freedom are intimately coupled in graphene, it is not clear *a priori* how a physical buckling or ripples might influence the electronic structure of topological insulators. Here we use Fourier transform scanning tunneling spectroscopy to determine the effects of a one-dimensional periodic buckling on the electronic properties of $Bi_2Te_3$. By tracking the spatial variations of the scattering vector of the interference patterns as well as features associated with bulk density of states, we show that the buckling creates a periodic potential modulation, which in turn modulates the surface and the bulk states. The strong correlation between the topographic ripples and electronic structure indicates that while doping alone is insufficient to create predetermined potential landscapes, creating ripples provides a path to controlling the potential seen by the Dirac electrons on a local scale. Such rippled features may be engineered by strain in thin films and may find use in future applications of topological insulators.

The helical Dirac fermions that exist on the surface of topological insulators (TIs) are condensed matter analogues of relativistic fermions extensively studied in high-energy physics[1, 2, 3, 4, 5, 6]. Many of the as yet theoretical scenarios in high-energy physics such as Majorana fermions and axion electrodynamics are also anticipated in TI[2, 7, 8, 9]. The greatest challenge to achieving these and other exotic states is the ability to control properties such as electron mobility or the position of chemical potential with respect to bands. Controlling and manipulating these properties on a local scale is particularly important in topological insulators where phase boundaries, dislocations and domain walls are predicted to play a special role. One possible method of engineering potential landscapes is via positioning dopants in a controlled fashion; however this approach has been shown to be impractical where previous studies have demonstrated no direct correlations between the spatial locations of the dopants and observed potential variations[10]. Extensive studies on the Dirac material graphene instead recommend another avenue to modulate electronic properties on a local scale via the creation of ripples.

Experimentally, 1D and 2D periodic ripples have been created in graphene by the strain arising from clamped edges or due to incommensurability with the substrate[11, 12]. Theoretical models of ripples in graphene suggest that the resultant local curvatures affect microscopic parameters such as the hopping matrix elements, which can change many aspects of the dispersion. In addition to changing microscopic parameters and creating local effects such as strong pseudo magnetic fields[13], ripples can result in an added periodic potential[14]. The effects of such periodic potentials on the electronic structure of graphene have been extensively explored theoretically[15, 16, 17, 18, 19] where it was shown that periodic potentials may not only change band parameters like the Fermi velocity but also induce additional zero energy modes and band gaps. However, while the charge and lattice degrees of freedom are intimately coupled in graphene, it is not clear *a* *priori* how a physical buckling such as ripples might influence the surface and bulk band structure in topological insulators. It is therefore essential to directly measure the spatially resolved electronic structure of topological insulators in the presence of surface ripples, especially in order to determine their viability in controlling local electronic properties.

In this work we investigate the effects of a one-dimensional periodic buckling and its associated chemical inhomogeneity on the electronic properties of the topological insulator, $Bi_2Te_3$[20]. We use low temperature scanning tunneling microscopy (STM) and spectroscopy, to probe the electronic structure of this surface. By carrying out local Fourier transforms of the interference patterns, we clearly demonstrate that in contrast to the non-local effects of chemical dopants, buckling has a well-defined, local effect on the Dirac dispersion. Our studies suggest that generating ripples provides an avenue to controllably engineer the potential landscape and sets the stage for future explorations of the effects of periodic potentials on the Dirac dispersion in topological insulators.

The single crystals of $Bi_2Te_3$ used in this study were grown via a standard modified Bridgman method[21]. The crystal surface was obtained by cleaving in UHV at room temperature before being inserted into the STM held at 4K. Annealed tungsten tips were used in this study, and conductance maps were obtained with a lock-in amplifier. Bulk transport measurements show that the sample is p-type. STM images of $Bi_2Te_3$ reveal two kids of regions: flat areas with no ripples as well as regions with a 1D periodic modulation i.e., stripes (Fig. 1(a)). The stripes have a periodicity of ~100 nm, ~0.1 nm height, and are aligned to within 5 degrees of ΓM (next-nearest neighbor) direction. Within one sample, the striped areas extend for at least a few micrometers and are adjacent to flat areas. Striped regions were observed on multiple samples with different tips. Based on the topographic characteristics alone, the stripes may originate from a modulation of the atomic positions (buckling) or purely electronic modulations due to charge or spin density wave order. The large ~ 100 nm periodicity and the lack of contrast reversal at opposite bias voltages (Fig. 1(c)) is however inconsistent with a conventional electronic charge density wave or a pure charge inhomogeneity, indicating that the stripes represent spatial modulations in height. While it is not explicitly clear which interactions[22] might produce these features, a probable explanation is that the buckling occurs during sample growth or cleaving due to strain. Strain for instance has been shown to result in a 1D reconstruction at various length scales and can affect both the lattice and the electronic structure[12, 23, 24, 25]. Empirically, in the present study, we expect that strain causes a modulation of the height of the atoms perpendicular to the surface, resulting in the periodic buckling observed in the topography.

We first discuss the influence of the buckling on the bulk bands. STM spectra of $Bi_2Te_3$ capture the density of states arising from the surface projection of the bulk bands as well as the Dirac surface states. A telltale feature in the spectroscopy is a broad resonance at higher energies, labeled E* in figure 1b, that has been previously identified as the energy of the bottom of the surface projected conduction band at the G point. Tracking the energy of E* allows us to follow spatial shifts in the bulk bands with respect to the Fermi energy. A plot of the energy of E* with position (E* map) as shown in figure 1d indicates that the E* values are modulated by the ripples. This correlation between the buckling and the bulk bands is also clearly observed in a 2D histogram of E* values plotted against relative height which reveals a bimodal split. On average, the value of E* is lower by 10 ± 2meV in the region surrounding the crest of the stripes (higher region) compared to the region around the troughs (Fig. 1e). These data indicate that the bulk bands in the crests are shifted in energy with respect to the bulk bands in the troughs by approximately 10meV.

We now investigate the effect of the ripples on Dirac surface state bands. Prior studies have shown that in samples with random potential variations the surface and bulk bands are shifted together[10] (rigid band shift) in response to local potentials, and the question is now whether the ripples cause a similar effect. To measure the effect of the stripes on the surface state, we obtained spatial differential conductance maps ($dI/dV(r,eV)$) as a function of energy (Fig. 2a). A 2D Fourier transform (FT) of the spatial map at a given energy captures the scattering vectors (q-vectors) that connect different points of the constant energy contours. This technique can be used to trace back to the original momenta (**k**) of the electrons and obtain the resulting dispersion. As shown previously[26,27,28] the dominant q-vector at higher energies in the high-symmetry direction is $q_{\Gamma M}$ as depicted in figure 2b. In order to capture the spatial variation of $q_{\Gamma M}$, we first carried out spatially resolved FTs of the conductance maps as shown in figure 2, which plots the spatially resolved q-vectors ($q_{\Gamma M}(r)$) and their averaged value (fig. 2g). These q-maps were extracted from peak positions along $\Gamma M$ of Fourier transforms of local (45nm*45nm sized areas corresponding to ~0.015 (1/nm) resolution in momentum space) patches. In contrast with previous studies showing no correlations between impurity density and electronic structure, our spatially resolved FTs reveal an unambiguous correlation between the topography with striped corrugation and the dispersion of the Dirac surface states. Specifically, the magnitude of $q_{\Gamma M}$ systematically shows a larger value at the crest of the stripes.

The clear correlation between striped topography and q-maps motivates us to define distinct larger regions for our Fourier transforms. Based on the topography, we separate the data into three regions with same coverage i.e., 1/3 of the total area (inset to figure 2a). Selective FFT of these larger regions gives us better q-resolution, allowing us to track the energy dependence of the q-vectors and thereby measure the relative energy shift of the surface state band. Since we do not observe any systematic anisotropies in q-vector magnitude along the three equivalent directions in k-space, we consider three $q_{\Gamma M}$'s equivalent and perform our analysis on the symmetrized Fourier transforms (FIg. 3b and c). We find that there is a measureable difference between the q-vectors obtained in the crests (red) and troughs (blue) regions (FIg. 3b, c and d). In converting back to the originating momenta we plot the surface state dispersion as shown in figure 3e. From this data we find that the surface state dispersion in the crests of the stripes is shifted to higher energies by 12 ± 2meV compared to that in the troughs of the stripes. In addition, Landau level measurements across the stripes show that the lower index (n=0 and 1) levels faithfully track the higher energy dispersion obtained from the interference patterns[29], proving that the electrons close to the Dirac point are equally affected by the stripes (Fig. 3e). Moreover, this energy scale is consistent with the shift obtained by us for the bulk bands from E* all of which indicate that the stripes create a one dimensional periodic potential landscape, which affects the bulk and surface state bands in a similar fashion.

Having determined that the stripes modulate both the bulk and surface bands, it is important to check the correlation between chemical variations due to impurity inhomogeneity and the topographic modulation in order to determine the role of the impurities if any in creating the observed periodic potential variations. Although previous studies of random doping induced inhomogeneity showed no spatial correlations between the dopant density and the potential variations, by modeling the impurities as a screened potential source, we find that there is a moderate, positive correlation between the topography and the potential variations resulting from one of the impurity species (labeled #3 in Fig. 4), which can be identified as an electron donor, with the other impurities showing far weaker or negligible correlations with the stripes (Fig. 4). This correlation coefficient (C(0)=0.3) however is far weaker than the much stronger correlation between the topography and electronic structure modulations as measured by the q-vector variations in the interference patterns as well as E* values (C(0)=0.65) (Fig. 5a), suggesting that the small changes in the impurity density may be a secondary effect. This is further supported by the equally strong on-site correlation between the topography and the n=1 Landau level energy (Fig. 5a) which directly reflects varying potential due to the stripes. Although this analysis does not exclude the possibility that disorder variations assist buckling, the analysis makes it clear that the potential variations originate from the ripple formation rather than changes in impurity density.

The picture that emerges from our data is the following. The periodic buckling represented by the stripes results in a smoothly varying periodic potential that is primarily responsible for trapping the carriers into the resultant 1D potential wells and shifting the bulk and surface states bands according to the rigid-band shift picture. An intriguing consequence of our observations is that in a magnetic field, such periodic chemical potential variations can play host to1D chiral metallic modes. Analogous to edge states in an integer quantum Hall system 1D modes are predicted at the spatial boundaries between two regions with successive half-integer fillings[2, 6, 30]. Such modes are interesting as practical realizations of 1D quantum wire and can be used to achieve dissipationless transport in one dimension (Fig. 5b). Finally our studies clearly show that for topological insulators, creating ripples is an efficient method for manipulating electronic properties on a local scale. We stress here that even though the ripples in these studies occur naturally from strain during sample growth or cleaving process, our results provide an essential first step to exploring the viability of strain engineering topological insulators. Such rippled features may be engineered for future devices by annealing and quenching of bulk crystals or by controllably inducing strain in thin films with suitable substrate choices or actively by piezo actuators.

Fig. 1

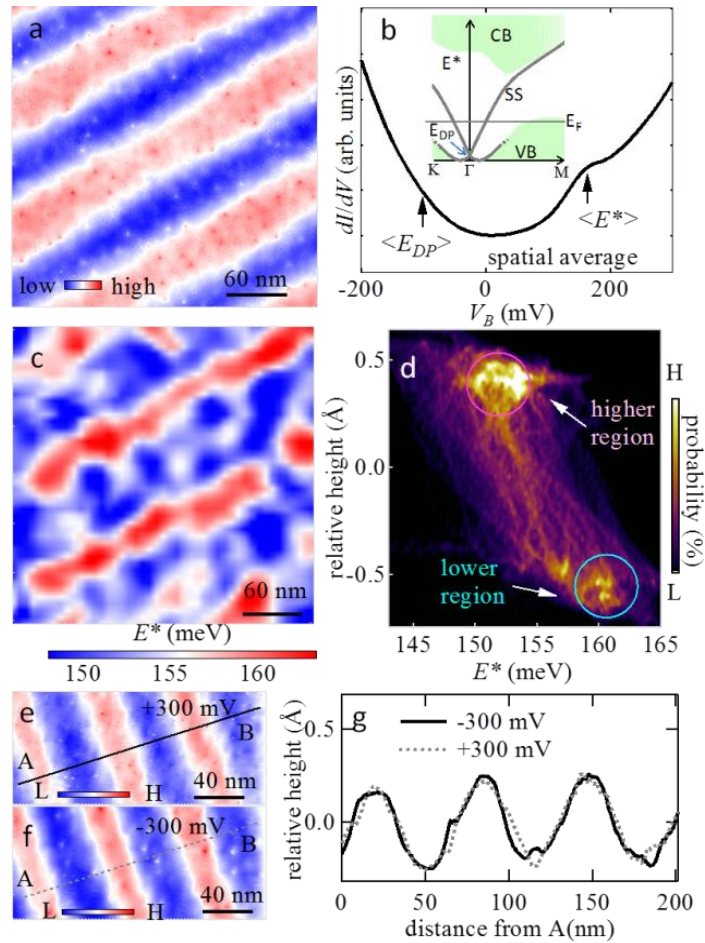

**Figure 1.** (a) Topography of stripes on the surface of $Bi_2Te_3$ obtained with bias voltage ($V_b$) =+250 (mV). (b) Spatially averaged *dI/dV* spectrum in the region shown in (a). The arrows show the approximate position for the Dirac point ($E_{DP}$) and the bulk conduction band edge around Γ point E*. The inset shows a schematic of the band structure of both surface and bulk bands together with Fermi energy $E_F$. (c) The spatial variation of E*. (d) Histogram of the locally obtained E* and the relative topographic height. (e) Topographic stripe and linecut across stripes (g) at $V_b$ =+300mV and -300mV respectively showing the absence of contrast reversal at positive (e) and negative (f) biases. These images were obtained on a different sample from that shown in (a)-(d).

Fig. 2

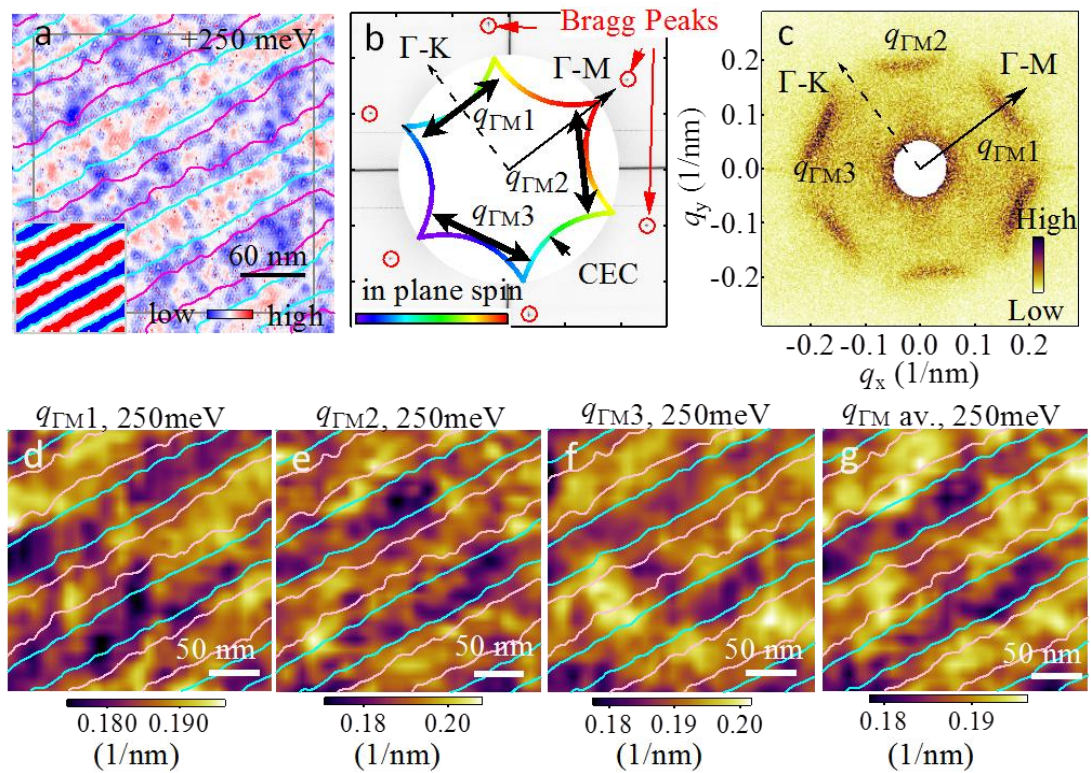

**Figure 2.** Spatial variations of q-vector along ΓM. (a) Conductance image at $V_b$= +250meV. The inset shows the definition of the three areas based on the height in the topography. The lines (red and blue) define the boundaries of these three regions. (b) FT image of topographic image shown in fig. 1(a). The inset shows a schematic constant energy contour (CEC) at +250meV (approximately 400 meV above the Dirac point), displaying a snow flake-like shape. (c) FT image of the conductance image (a). Three characteristic q-vectors $q_{ΓM}1$, $q_{ΓM}2$, and $q_{ΓM}3$ connect the well-nested portions of the CECs shown schematically in (b). Spatial maps of $q_{ΓM}1$, $q_{ΓM}2$, $q_{ΓM}3$, (d)-(f), and averaged value (g) for the area enclosed by the grey square shown in (a).

Fig. 3

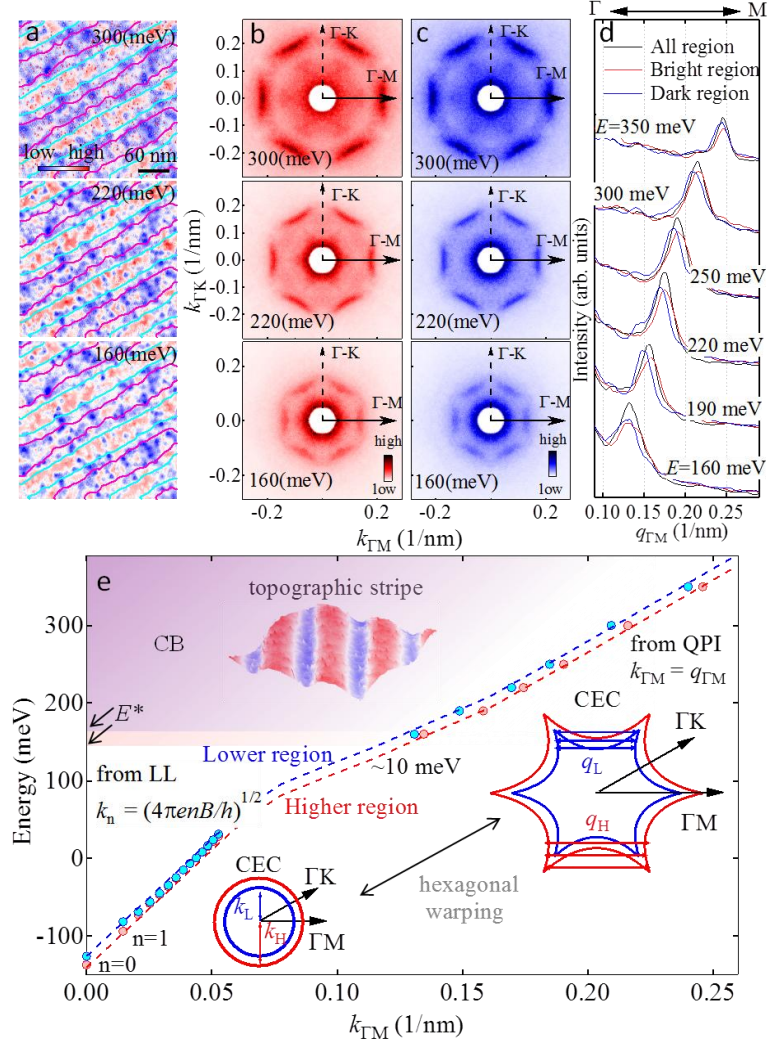

**Figure 3.** (a) *dI/dV (r, eV)* (conductance) maps at +300meV, +220meV, and +160meV (top to bottom), with the lines defining the threee regions shown in the inset of fig. 2(a). FT images for the red (b) and blue (c) regions, respectively. The FTs are 6-fold symmetrized and have been rotated such that the ΓM direction is along to x-axis. (d) Line cut along ΓM of FT images of the two regions, and their energy dependence. (e) Peak positions obtained from line cuts shown in (d) together with data points from from previously reported Landau level (LL) spectra at B=7T[29]. The insets show schematic constant energy contours for the red and blue regions based on our data.

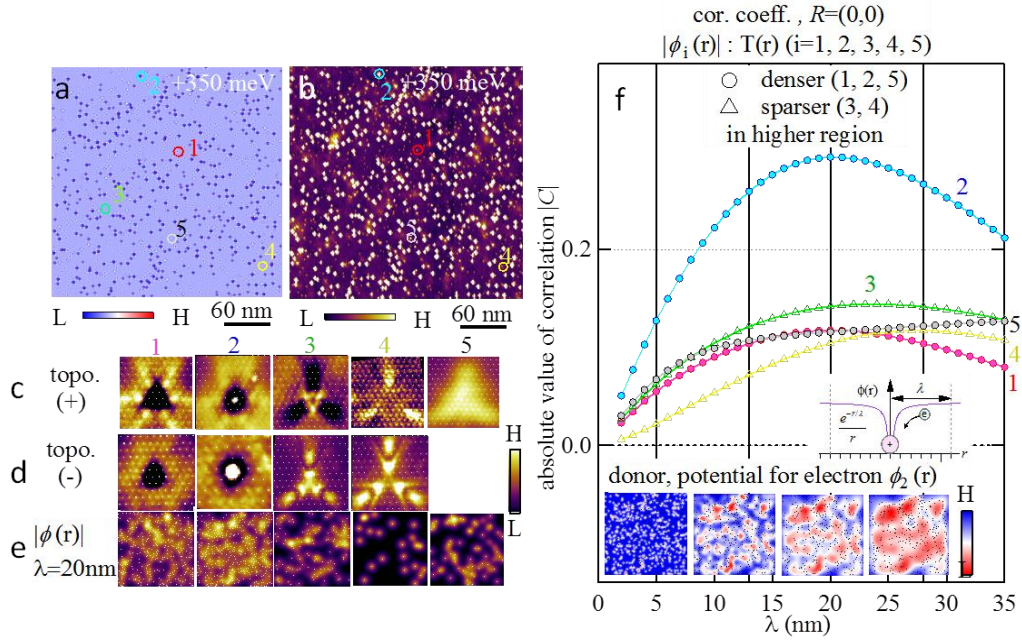

**Figure 4.** (a) Topography of the striped region (stripes filtered) and (b) conductance map at +350 mV in same area as Fig. 1(a) along with a catalog of the five clearly visible impurities (c)-(e) in $Bi_2Te_3$. (c) and (d) show high resolution topographies at positive and negative bias, respectively of the five primary impurities identified in a. The first two show a clear halo feature and stronger unoccupied conductance as seen in (a), which suggests that they act as electron donors. (e) Variation of the absolute value of potential $|\phi_i(r)|$ at a chosen value of $\lambda=20$ nm. The definition of the length scale $\lambda$ is shown in the inset to (f), where $\phi_i(r)$ is a simple model for the effective potential generated by the impuritiy. (f) Quantitative evaluation of the correlation between $|\phi_i(r)|$ and topographic stripes T(r). The graph shows the evolution of the absolute value of the onsite correlation coefficients, $|C[|\phi_i(r)|: T(r+R)]|$ at R=0, for the various impurities as a function of the screening length $\lambda$. As seen in the graph as well as visually in the inset, the sptial extent of the potential as determined by $\lambda$ influences the calculated correlation coefficient.

Fig. 5

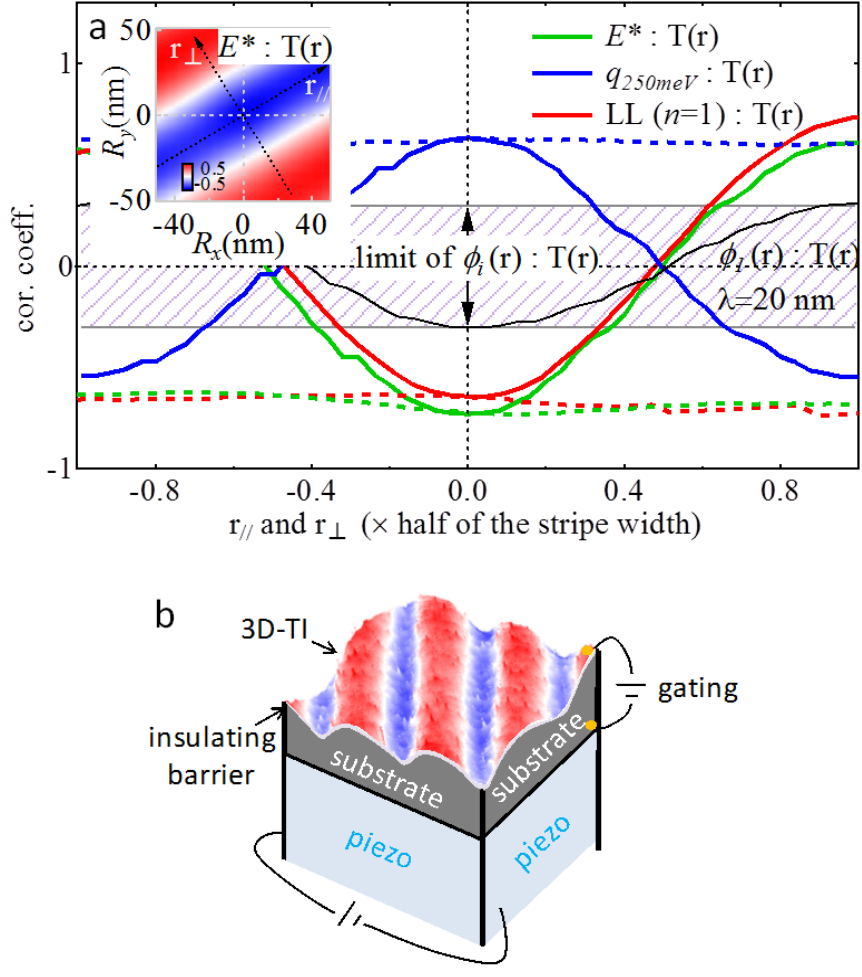

**Figure 5.** (a) Plot of the correlation coefficient C[ $E_{n=1}$(r): T(r+R)] (red), C[ $q_{\Gamma M}$(r): T(r+R)] (blue), and C[ $E^*$(r): T(r+R)] (green) along a line parallel (dashed line) and perpendicular (solid line) to the stripe direction. The maximum limit of the correlation coefficient between $\phi_i$(r) and T(r) (0.3) is depicted by the shaded area. The inset image shows an image plot of C[ $E^*$(r): T(r+R)] for various R. (b) Schmatic of a possible stripe based device with the stripes being generated by the strain caused by a piezoelectric crystal.